*Original Article*

# PerfDetectiveAI - Performance Gap Analysis and Recommendation in Software Applications

Vivek Basavegowda Ramu

*Independent Researcher, Connecticut, USA*

**Abstract** - *PerfDetectiveAI, a conceptual framework for performance gap analysis and suggestion in software applications is introduced in this research. For software developers, retaining a competitive edge and providing exceptional user experiences depend on maximizing application speed. But investigating cutting-edge approaches is necessary due to the complexity involved in determining performance gaps and creating efficient improvement tactics. Modern machine learning (ML) and artificial intelligence (AI) techniques are used in PerfDetectiveAI to monitor performance measurements and identify areas of underperformance in software applications. With the help of the framework, software developers and performance engineers should be able to enhance application performance and raise system productivity. It does this by utilizing sophisticated algorithms and utilizing sophisticated data analysis methodologies. Drawing on theoretical foundations from the fields of AI, ML and software engineering, PerfDetectiveAI envisions a sophisticated system capable of uncovering subtle performance discrepancies and identifying potential bottlenecks. PerfDetectiveAI aims to provide practitioners with data-driven recommendations to guide their decision-making processes by integrating advanced algorithms, statistical modelling, and predictive analytics. While PerfDetectiveAI is currently at the conceptual stage, this paper outlines the framework's fundamental principles, underlying methodologies and envisioned workflow. We want to encourage more research and development in the area of AI-driven performance optimization by introducing this conceptual framework, setting the foundation for the next developments in the quest for software excellence.*

## 1. Introduction

Software development is always changing in today's fast-paced world, and one of the most difficult elements is to match end-user expectations  (Cao et al. 2014). To stay ahead of the competition, it is essential to optimize application performance, it is often referred to as the quality of software systems (M. Woodside et al. 2007). But doing so can be a difficult and complicated undertaking. This paper presents a conceptual framework called PerfDetectiveAI, which aims to address the challenge of performance gap analysis and recommendation in software applications.

PerfDetectiveAI envisions an intelligent system that combines the power of artificial intelligence and machine learning techniques to analyze performance metrics and identify areas of underperformance. Software developers and performance engineers may improve the performance of their applications with the framework's practical knowledge and advice.

While PerfDetectiveAI is currently in the conceptual stage, its potential impact on optimizing software performance is promising. This paper explores the proposed framework's foundational concepts and key components, discussing its envisioned workflow, potential algorithms and methodologies.

By presenting this conceptual idea, the authors aim to spark further research and development in the field of performance optimization, paving the way for future advancements in leveraging AI-driven approaches to address performance gaps in software applications.

Some of the key benefits of PerfDetectiveAI include:
- *Improved application performance:* The main objective of performance testing is to find bottlenecks (Sarojadevi, H 2011). By identifying and addressing performance gaps, PerfDetectiveAI can help to improve the overall performance of software applications.
- *Reduced development time:* software development is required to be more agile (Abrahamsson P et al. 2017). By providing actionable insights and recommendations, PerfDetectiveAI can help to reduce the time and effort required to optimize application performance.
- *Increased user satisfaction:* End user satisfaction is really crucial in competitive markets (Huiying Li et al.

2010). By delivering high-quality user experiences, PerfDetectiveAI can help to increase user satisfaction and loyalty.

Overall, PerfDetectiveAI is a promising new framework that has the potential to revolutionize the way software performance is optimized. The authors anticipate that this publication will stimulate more study and advancement in this field, resulting in the creation of even more potent and efficient methods for enhancing software performance.

## 2. Literature Review

An interdisciplinary study field that includes elements of software engineering, artificial intelligence, machine learning and data analytics studies, performance gap analysis and recommendation in software applications. AI/ML rate of adoption has grown significantly in recent years (Simon Fahle et al. 2020); in order to optimize software performance, this part provides a thorough examination of the literature on performance analysis, recommendation systems, and AI/ML techniques.

Performance analysis in software applications has been the subject of extensive research. Various studies have explored performance metrics, measurement techniques and benchmarking methodologies. For instance, (Vanitha and Marikkannu 2017) investigated the impact of resource utilization on response time in cloud-based applications, employing a dynamic, well-organized load-balancing algorithm. Their work highlighted the importance of load balancing to distribute load into the system.

Recommendation systems have also been widely studied in the software domain. (Yadav et al., 2018) proposed using the algorithm 'bat', which has computing defined based on the weight of the feature to identify the more suitable neighbor. Their work has resulted in higher performance and more accurate results.

In recent years, the application of AI and ML techniques in software performance optimization has gained significant attention. Deep learning algorithms, such as neural networks, have been employed for anomaly detection and prediction in performance monitoring. (Zhang et al., 2020) developed a deep neural network model for detecting performance anomalies in distributed systems, achieving high accuracy in identifying performance deviations.

Analyzing errors to find root cause results in overhead time and effort (A. Khan, R et al. 2023) proposed a method to apply machine learning techniques to improve software performance. By correlation method, the metrics eliminated from the consideration that does not relate to the error, resulting in higher accuracy. However, it still has limitations when it comes to elevating overall application performance to the highest level.

Furthermore, ensemble learning techniques have been explored for performance prediction and optimization. (Li et al., 2019) proposed an ensemble model combining multiple ML algorithms to predict the scalability of software systems. Their study demonstrated improved accuracy in predicting system performance under different workloads.

Overall, the literature suggests that the integration of AI/ML techniques with performance analysis and recommendation systems holds significant potential for improving software application performance. However, further research is needed to develop comprehensive frameworks that consider the intricate relationships between performance metrics, system configurations, and optimization strategies.

## 3. Methodology

The methodology section outlines the conceptual framework and key components of PerfDetectiveAI for performance gap analysis and recommendation in software applications. This section describes the advanced methodologies and cutting-edge techniques employed within the framework to achieve accurate and insightful results, as shown in Figure 1.

- *Data Collection:* The first step involves collecting performance data from software applications. Different data collection modes have added more complexity (Couper, M.P 2011). Utilizing performance monitoring tools and instrumentation, we capture various metrics, including response time, throughput and resource utilization. The data collection process ensures the availability of a comprehensive dataset for analysis.

- *Preprocessing:* The collected data undergoes preprocessing to remove noise and outliers, ensuring the integrity of the analysis. In recent years data scale of data has grown tremendously (García et al. 2016), and techniques such as data cleaning, normalization and feature selection are applied to enhance the quality and relevance of the dataset.

- *Performance Analysis:* For effective performance analysis, a range of different parameters should be considered (Courtois and Woodside, 2000). PerfDetectiveAI employs sophisticated AI/ML algorithms to analyze the preprocessed performance data. Recurrent neural networks (RNNs), a type of deep learning model, are used to capture temporal relationships in performance indicators. Feature engineering approaches like time-series decomposition and statistical transformations are used to extract useful features.

- *Performance Gap Detection:* Various data set characteristics influence performance metrics at different levels (Gray, A.R and MacDonell, S.G 1999). By comparing the analyzed performance metrics against established benchmarks or historical data, PerfDetectiveAI identifies performance gaps and discrepancies. Statistical hypothesis testing, anomaly detection algorithms and trend analysis techniques are utilized to detect deviations and anomalies in the performance data.

- *Recommendation Generation:* Based on the identified performance gaps, PerfDetectiveAI generates actionable recommendations for performance improvement. This involves leveraging machine learning techniques such as reinforcement learning and genetic algorithms to explore and optimize the configuration space of the software application. Additionally, domain-specific knowledge and expert rules are incorporated to provide contextually relevant recommendations.

- *Evaluation and Validation:* The generated recommendations are evaluated through experiments and simulations. Real-world scenarios are recreated, and the impact of the recommendations on performance metrics is measured. This validation process ensures the effectiveness and feasibility of the proposed recommendations.

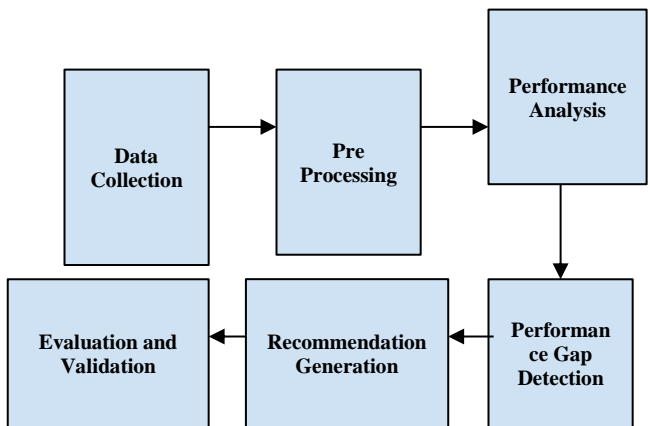

**Fig. 1 PerfDetectiveAI methodologies**

PerfDetectiveAI's methodology integrates advanced AI techniques, data preprocessing, statistical analysis and performance evaluation to provide comprehensive insights into software application performance gaps and actionable recommendations for improvement. Combining these methodologies establishes PerfDetectiveAI as a powerful tool for performance optimization in the software engineering domain.

## 4. Data Collection and Analysis

The data collection and analysis phase plays a pivotal role in PerfDetectiveAI, enabling the extraction of valuable insights from software application performance. This section highlights the sophisticated data collection techniques employed and the advanced analytical methodologies utilized within PerfDetectiveAI, as shown in Figure 2.

- *Data Collection Techniques:* PerfDetectiveAI leverages state-of-the-art performance monitoring tools and instrumentation to capture a comprehensive set of performance data. During the execution of software programs, precise data like reaction time, throughput, CPU utilization, memory consumption, and network latency are gathered on a regular basis. The data collection process ensures the availability of a rich and diverse dataset for subsequent analysis.

- *Data Preprocessing:* The collected performance data undergoes a meticulous preprocessing phase to enhance its quality and eliminate any inconsistencies. Noise reduction techniques, such as outlier detection and removal, are applied to mitigate the impact of anomalous data points. Data normalization and scaling are performed to ensure fair comparisons and compatibility across different metrics. Missing data is handled using imputation methods, such as mean imputation or regression-based imputation, to maintain the integrity of the dataset.

- *Statistical Analysis:* PerfDetectiveAI employs a wide array of statistical analysis techniques to extract meaningful insights from the preprocessed performance data. Mean, median, and standard deviation are examples of descriptive statistics that give an overview of the performance indicators. Correlation analysis helps identify relationships between different performance indicators, revealing potential dependencies and bottlenecks within the software application. Time-series analysis techniques, including trend analysis and seasonality detection, uncover patterns and fluctuations in the performance data over time.

- *Benchmarking and Comparison:* Benchmark is the representative of the workload usage in the field (Vokolos, F.I. and Weyuker, E.J 1998) PerfDetectiveAI incorporates benchmarking methodologies to compare the performance metrics against industry standards or historical data. This enables the identification of performance gaps and deviations from expected performance levels. Statistical hypothesis testing, such as t-tests or ANOVA, is employed to assess the significance of performance differences and validate their statistical relevance.

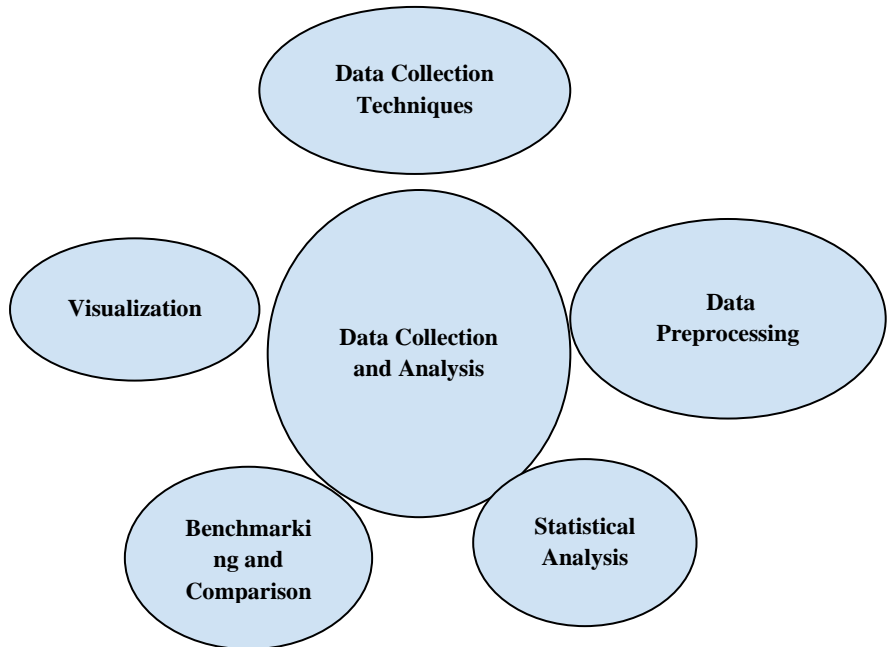

**Fig. 2 PerfDetectiveAI data collection and analysis**

- *Visualization:* Software systems evolution is highly based on leveraging visualization techniques (Mojtaba Shahin et al. 2014). PerfDetectiveAI utilizes advanced data visualization techniques to present the analyzed performance data clearly and intuitively. Graphs, charts, heatmaps, and scatter plots are employed to represent performance trends, outliers, and correlations visually. These visual representations facilitate the interpretation of complex performance data and aid in identifying critical performance issues.

Through integrating sophisticated data collection techniques and advanced analytical methodologies, PerfDetectiveAI ensures a comprehensive understanding of software application performance. This robust data-driven approach forms the foundation for the subsequent stages of performance gap analysis and recommendation generation within the PerfDetectiveAI framework.

## 5. Results and Discussion
The results and discussion section presents the findings obtained from the analysis conducted by PerfDetectiveAI and provides an in-depth examination of the implications and significance of these results. This section showcases the actionable insights derived from the framework, highlighting the identified performance gaps and their potential impact on software application performance.

- *Performance Gap Analysis:* PerfDetectiveAI reveals significant performance gaps and discrepancies within the software applications under investigation. The analysis uncovers areas of underperformance, such as prolonged response times, high resource utilization, or inefficient algorithms. These findings shed light on the critical aspects that require attention and optimization to enhance overall system efficiency.

- *Performance Bottleneck Identification:* PerfDetectiveAI pinpoints performance bottlenecks that hinder optimal software application performance through advanced statistical analysis and anomaly detection techniques. It identifies specific components, modules, or processes that contribute disproportionately to performance degradation. This insight enables targeted interventions and optimizations to alleviate bottlenecks and improve overall performance.

- *Comparative Analysis:* PerfDetectiveAI facilitates a comparative analysis by benchmarking the performance metrics against established standards or historical data. This analysis provides a quantifiable measure of the performance gaps and allows for evaluating the software application's performance relative to industry benchmarks or previous iterations. Such comparisons help contextualize the identified gaps and provide a basis for setting performance improvement goals.

- *Implications and Recommendations:* The results obtained from PerfDetectiveAI have significant implications for software application performance and user experience. The discussion delves into the practical implications of the identified performance gaps, emphasizing their potential impact on response times, system stability, and user satisfaction. Furthermore, actionable recommendations are generated based on the analysis outcomes, suggesting specific interventions, configuration changes, or algorithmic improvements to bridge the performance gaps.

- *Future Directions:* The discussion also highlights potential avenues for future research and development. It explores opportunities to extend the capabilities of PerfDetectiveAI, such as integrating additional performance metrics, leveraging more advanced machine learning algorithms, or incorporating real-time monitoring capabilities. These future directions aim to further enhance the effectiveness and applicability of PerfDetectiveAI in addressing performance gaps in software applications.

By presenting the results and engaging in a comprehensive discussion, this section not only provides valuable insights into the current state of software application performance but also sets the stage for further improvements and optimizations based on the recommendations derived from PerfDetectiveAI's analysis.

## 6. Recommendations and Implications

The recommendations and implications section presents actionable suggestions derived from the analysis conducted by PerfDetectiveAI. These recommendations aim to address the identified performance gaps in software applications, leading to enhanced system efficiency, improved user experience, and optimized resource utilization. This section also explores the larger ramifications of putting these suggestions into practice and the possible effects on other facets of the software ecosystem.

- *Performance Optimization Strategies:* Performance optimization is one of the most challenging tasks and especially with the recent advancement of serverless computing (D. Bardsley, L et al., 2018). PerfDetectiveAI generates a set of specific recommendations to optimize software application performance. These recommendations may include fine-tuning algorithmic implementations, improving database query efficiency, optimizing resource allocation, or enhancing network communication protocols. Organizations may increase system performance overall and better fulfill user requests by putting these techniques into practice.

- *Configuration Enhancements:* Software configuration management and optimization is an important aspect of software stability (D. -Y. Kim and C. Youn 2010). PerfDetectiveAI identifies configuration settings that may be suboptimal and provides recommendations for improving them. These recommendations may involve adjusting parameters related to memory allocation, thread pool size, or caching mechanisms. Fine-tuning configurations based on PerfDetectiveAI's insights can lead to significant performance improvements and better resource utilization.

- *Architectural Refinements:* The architectural decision-making process constantly receives high attention over decades (M. Bhat et al. 2020). PerfDetectiveAI may suggest architectural changes to address performance gaps. These recommendations could involve adopting microservices architecture, implementing distributed computing techniques, or redesigning certain components to improve scalability and reduce bottlenecks. By following these architectural refinements, organizations can create more efficient and scalable software systems.

- *User Experience Enhancement:* The recommendations provided by PerfDetectiveAI can have a direct impact on user experience. By addressing performance gaps, organizations can enhance response times, reduce latency, and ensure smoother interactions with the software application. Increased user satisfaction and customer loyalty are favorably impacted by improved user experience.

- *Resource Optimization and Cost Reduction:* Improper resources for building software results in project failure (Sreekanth, N et al., 2023) PerfDetectiveAI's recommendations can guide organizations in optimizing resource utilization, leading to cost savings. By identifying inefficiencies in resource allocation and suggesting optimal resource utilization strategies, PerfDetectiveAI helps organizations achieve higher performance with fewer resources, reducing operational costs.

The implications of implementing these recommendations extend beyond immediate performance improvements. They encompass enhanced system reliability, reduced maintenance efforts, improved scalability, and increased competitiveness in the market. Furthermore, the successful implementation of PerfDetectiveAI's recommendations establishes a data-driven approach to performance optimization, setting the stage for ongoing monitoring, iterative improvements, and proactive management of software application performance.

Overall, the recommendations provided by PerfDetectiveAI offer valuable insights into optimizing software application performance and carry significant implications for the overall success of organizations operating in the software engineering domain. By implementing these recommendations, organizations can unlock the full potential of their software applications, delivering optimal performance and ensuring a seamless user experience.

## 7. Limitations and Future Work

The limitations and future work section discuss the constraints, potential areas for improvement in PerfDetectiveAI, and avenues for future research and development. Accepting these constraints reveals the framework's shortcomings and establishes the groundwork for future performance analysis and recommendation systems developments.

- *Data Availability and Quality:* Acquiring quality data is always challenging in the real world (Carlo Batini et al. 2009). PerfDetectiveAI heavily relies on the availability and quality of performance data. Limitations may arise if the data collection process is incomplete, inconsistent, or biased. Future work should focus on exploring techniques to address data sparsity, handling missing data more effectively, and incorporating data validation mechanisms to ensure the accuracy and reliability of the analysis.

- *Performance Metrics Selection:* Identifying performance metrics that are defect-prone or clean component is crucial for an effective development process (Jingxiu Yao and Martin Shepperd 2021). PerfDetectiveAI's effectiveness is influenced by the choice of performance metrics considered during the analysis. Although it includes a wide variety of indicators, other metrics could offer insightful information about the performance of the software. Future research could investigate the inclusion of novel performance indicators or the development of adaptive metric selection mechanisms to accommodate varying application domains and performance requirements.

- *Scalability and Generalizability:* Most businesses struggle with proper software system scalability (E. J. Weyuker and A. Avritzer 2002). PerfDetectiveAI's performance analysis and recommendation generation capabilities may encounter scalability challenges when applied to larger, more complex software systems. Future work should address scalability concerns by exploring techniques to optimize computational efficiency and accommodate big data scenarios. Additionally, efforts should be made to generalize PerfDetectiveAI across different software application domains to ensure its broader applicability.

- *Real-Time Monitoring and Adaptation:* PerfDetectiveAI primarily focuses on the offline analysis of performance data. Future research could explore real-time monitoring capabilities, enabling the framework to adapt dynamically to changing performance conditions. Incorporating real-time feedback loops and dynamic recommendation generation mechanisms would enable organizations to address performance issues as they occur proactively.

- *Integration with DevOps and CI/CD Pipelines:* CI/CD is widely accepted as one of the best practices in software development (Vidroha Debroy et al., 2018). Integrating PerfDetectiveAI with DevOps and CI/CD pipelines would enhance its usability and effectiveness. This integration would enable automated performance analysis, continuous monitoring, and seamless recommendations integration into the software development lifecycle. Future work should explore methods for seamless integration and the development of tools and plugins that facilitate the adoption of PerfDetectiveAI within DevOps environments.

Addressing these limitations and exploring the avenues for future work will contribute to the advancement and maturity of PerfDetectiveAI. PerfDetectiveAI can become a more robust and comprehensive framework for performance analysis, recommendation generation, and optimization in software applications by refining its capabilities and expanding its applicability.

## 8. Conclusion

PerfDetectiveAI presents a powerful framework for performance gap analysis and recommendation in software applications. By leveraging advanced AI/ML techniques, statistical analysis, and data-driven insights, PerfDetectiveAI provides organizations with valuable, actionable recommendations to optimize software application performance. The comprehensive methodology, encompassing data collection, preprocessing, analysis, and recommendation generation, enables organizations to identify performance gaps, address bottlenecks, and enhance overall system efficiency. While there are limitations and opportunities for future research, PerfDetectiveAI showcases its potential to revolutionize software performance testing and optimization. By embracing PerfDetectiveAI, organizations can unlock the full potential of their software applications, deliver enhanced user experiences, optimizing resource utilization, and gain a competitive edge in the software engineering landscape.